\documentclass[11pt,letterpaper,notoc]{JHEP3}
\usepackage{graphicx}
\usepackage{amsfonts}
\usepackage{amsmath}
\usepackage{epsfig}
\newcommand{\SM}{Standard Model}
\title{Neutrino Masses in Composite Little Higgs Model}
\author{Jaeyong~Lee\\
Department of Physics, Box 1560,
University of Washington, Seattle, WA 98195-1560 \\
E-mail: \email{jaeyong@u.washington.edu}}
\preprint{UW/PT-05-01}

\abstract{The composite little Higgs model, a UV completion 
for the $SU(5)/SO(5)$ little Higgs model, incorporates supersymmetry into
strong gauge dynamics. We extend the study of  flavor physics in the model,
and find that it is similar to the bosonic technicolor model.
Lepton flavor violations and neutrino mass matrix arise
once $R$-parity violating superpotential is introduced to the model,
as in the MSSM.
We identify various low-energy effective $\Delta L=2$ lepton flavor
violating operators, and find that most of them are similar to those 
of the $R$-parity violating MSSM. There is a new operator which
involves only leptons and the pseudo-Nambu Goldstone bosons of the little
Higgs model. We further study a possibility that this operator gives
a dominant contribution to the neutrino mass matrix.}
\begin{document}

\section{Introduction}

In the \SM~(SM) there are four particles whose masses are still unknown,
the Higgs boson, $h$ and the three neutrinos ($\nu_e, \nu_\mu$,$\nu_\tau$). 
The direct searches at LEP set the lower bound on the SM Higgs mass,
$\sim$114 GeV \cite{Barate:2003sz} while precision electroweak data,
obtained at the colliders (LEP, SLD, C0, CDF) in the framework of the SM,
suggest the SM Higgs boson must be relatively light and its upper
limit is $\sim$210 GeV. This gives us a hope that the SM Higgs is
light enough to be discovered at the LHC in near future.
On the other hand, through the past few years, several neutrino experiments
have identified convincingly that neutrinos are indeed massive and
like the quarks, mix among themselves \cite{Giacomelli:2004va}.
This is a solid evidence of the new physics beyond the SM. 

There are several popular mechanisms for understanding small neutrino
masses. In a generic see-saw mechanism, very massive right-handed
neutrinos are integrated out at high energy and neutrino mass is inversely
proportional to the mass scale of the right-handed neutrino, $M_R$ and
is quadratic in the Higgs {\it vev}, $m_\nu\sim v^2/M_R$ where $v$ is the
{\it vev} of the SM Higgs.
On the contrary,  triplet scalar models admit a neutrino mass which is
linear to the {\it vev} of the Higgs triplet, $m_\nu\sim y v'$ where $y$ is
a dimensionless coupling constant \cite{Chun:2003ej}.
As a third way, in supersymmetric theories a tiny neutrino mass
\cite{Hall:1983id} is induced by $R$-parity violating (RPV) superpotential. 

In all these mechansims, the relevant Higgs are assumed to be
fundamental particles with masses. But there is a different possibility 
that the Higgs is a pseudo-Nambu-Goldstone boson (pNGB).
Three years ago, Arkani-Hamed, Cohen, Katz, and Nelson constructed
the successful Higgs model with pNGBs, called the
``the Littlest Higgs model" \cite{Arkani-Hamed:2002qy,Han:2003wu}.
In the Littlest Higgs model, the SM Higgs doublet belongs to a set of
pNGBs in a global $SU(5)/SO(5)$ nonlinear sigmal model, and
other elements of the pNGBs are grouped into a complex Higgs triplet.
The nonlinear transformation of the pNGBs under the collective global
symmetries naturally ensures the absence of the SM Higgs mass term
of the form $m^2|h|^2$. However, at the electroweak scale the SM Higgs
potential is induced by the Coleman-Weinberg potential
\cite{Coleman:1973jx} in the gauge sector as well as in the fermion sector,
and the SM Higgs mass is generated by loop contributions from the
massive particles.

In the present article, we focus on neutrino physics in the $SU(5)/SO(5)$
little Higgs model. To account for small neutrino masses, the pNGBs interact
with neutrinos. But it is a drawback that the UV cutoff of the little Higgs
models is relatively low, being typically in the range of 10 TeV
$\lesssim\Lambda\lesssim$ 100 TeV. Thus we suggest that there
is an unknown symmetry that almost forbids the couplings
between the pNGBs and neutrinos in a UV completion of the little
Higgd models. We search for the mechanism which introduces small
neutrino masses. So far there have been two known UV completions
of the little Higgs models:
one is the composite little Higgs model (CLHM) \cite{Katz:2003sn}, 
and the other is suggested by  the ADS/CFT correspondence
\cite{Thaler:2005en}. Both models are based on the gauge group
$SU(2)_1\times SU(2)_2\times U(1)_Y$, and need strong dyamics
at the TeV scale. Here we further investigate the CLHM to account
for small neutrino masses.

The outline of the rest of the paper is as follows: In section 2 we
review the Higgs sector of the $SU(5)/SO(5)$ little Higgs model, and
take into account the low-energy effective $\Delta L=2$ lepton flavor 
violating (LFV) operator. In section 3 we review the composite little
Higgs model, and  further investigate flavor physics
in the composite little Higgs model. In section 4 we introduce
$R$-parity violation (RPV) in the composite little Higgs model,
and then identity various $\Delta L=2$ LFV operators.
In section 5, we investigate neutrino mass matrix from the LFV
operators coming from PRV couplings. In section 6, we specify
the case where the LFV operator, coupled only  to the pNGB,
gives a dominant neutrio mass matrix.
Finally, we draw a conclusion in section 7.
 
\section{$SU(5)/SO(5)$ Little Higgs Model}

The Littlest Higgs Model begins with a global $SU(5)$ symmetry,
with a locally gauged subgroup
$[SU(2)_1\times U(1)_1]\times[SU(2)_2\times U(1)_2]$
\cite{Arkani-Hamed:2002qy}.
The $SU(5)$ global symmetry breaking is spontaneously broken down to
its subgroup $SO(5)$ at the scale $f\sim\,1\,$ TeV resulting in fourteen NGBs.
This breaking arises from a {\it vev} of the $5\times 5$ symmetrical matrix
$\Xi$, which transforms $\Xi\rightarrow V\Xi V^T$ under $SU(5)$.
\begin{equation}
\Sigma_0=\left(\begin{array}{ccc}
 & & {\mathbf 1}_{2\times2} \\
 & 1 &  \\
 {\mathbf 1}_{2\times2} & & \end{array}\right).
 \end{equation}
The four NGBs are eaten by the gauge bosons so that the gauge
group $[SU(2)\times U(1)]^2$ is broken down to the electroweak
gauge group $SU(2)_L\times U(1)_Y$.
The remaining ten NGBs can be parameterized by the non-linear
$\Sigma$ field
\begin{equation}
\Sigma=e^{i\Pi/f}\Sigma_0 e^{i\Pi^T/f} 
=e^{2i\Pi/f}\Sigma_0,\qquad
\Pi=\left( \begin{array}{ccc}
 & h^\dagger\!/\sqrt{2} & \phi^\dagger \\
h/\sqrt{2} & & h^\ast\!/\sqrt{2} \\
\phi & h^T\!/\sqrt{2} & \end{array} \right), \label{eq:Sigma}
\end{equation}
and consist of $h=(2,1/2)$ and $\phi=(3,1)$ in $SU(2)_L\times U(1)_Y$: 
\begin{equation}
h = \left(\begin{array}{cc} h^+ &  h^0 \end{array} \right)\,, 
\quad
\phi=\left( \begin{array}{cc} \phi^{++} & \phi^+\!/\sqrt{2} 
\\ \phi^+\!/\sqrt{2} & \phi^0 \end{array} \right).
\end{equation}

The $h$ is identified as the SM Higgs while the $\phi$ is an addition
to the SM. The mechanism of collective symmetry breaking generates the 
potential for $\phi$ and $h$: $\phi$ acquires
a large mass of order $gf$ from one-loop gauge quadratic 
divergent part of the Coleman-Weinberg potential, while
$h$ acquires a negative mass squared parameter 
from logarithmically enhanced parts of the Coleman-Weinberg
potential in the gauge sector, in the scalar sector and in the fermion
(third generation up-type quark) sectors. The low energy effective
potential admits {\it vev}'s for $h$ and $\phi$:
$\langle h^0\rangle=v/\sqrt{2}$, and $\langle \phi^0\rangle=-iv'$
\footnote{From contraint on $\rho$ parameter, $v'\ll v$}.

The littlest Higgs model has an ambiguity of two 
$U(1)$'s charge assignements in the fermion sector \cite{Han:2003wu}.
To avoid this ambiguity and an extra heavy $U(1)$ gauge field at low
energy, one can consider a simpler model, in which only  one 
$U(1)$ is gauged. This makes it easier to build a UV
completion of a $SU(5)/SO(5)$ little Higgs model. In the following,
we consider the  $SU(5)/SO(5)$ little Higgs model with the 
$SU(2)_1\times SU(2)_2\times U(1)_Y$ gauge group. The ungauged field,
corresponding to the extra $U(1)$ gauge field in the littlest Higgs model,
remains an exact NGB to the order in which we are working. The impact
of this axion-like field at the electroweak scale is disccussed in 
Ref. \cite{Kilian:2004pp}. We neglect this axion-like field in the following.

We now take into account the interactions between $\Sigma$ and the SM leptons.
The SM lepton doublet and singlet are transformed as $\ell=(2,1)$
and $e^c=(1,1)$ in the $SU(2)_1\times SU(2)_2$ gauge group.
To account for mass of the charged leptons,
the LH model contains the Yukawa interaction of the form
\cite{Arkani-Hamed:2002qy}
\begin{equation}
{\cal L}_{Yuk}=
\frac{1}{2}\lambda^e_{\alpha\beta} f \epsilon^{ijk} \epsilon^{xy} (\ell_i)^\alpha
\Sigma^\ast_{jx} \Sigma^\ast_{ky} (e^c)^\beta + h.c.\label{eq:lepmass}
\end{equation}
where $\alpha,\beta$ represents the generation and
$i,j,k,x,y$ represent the component in the $SU(5)$ representation.
Note that $i,j$ are summed over 1,2 (index in the gauge group
$SU(2)_1$), $k=3$, and $x,y$ are summed over $4,5$.
As we expand (\ref{eq:lepmass}) in power of $1/f$, we read the SM
Yukawa couplings at leading order.

There may be other interactions between $\Sigma$ and the SM leptons
which admit lepton flavor violations to account for tiny neutrino
masses observed in neutrino oscillation experiments.
Provided that neutrino mass is induced primarily from
$\Sigma$ field,  one can consider a LFV operator of the form
\begin{equation}
{\cal L}_{LFV}=
 z_{\alpha\beta}\epsilon^{ij}\epsilon^{kl}f (\bar {\ell^c_i})^\alpha
\Sigma^\ast_{jk} (\ell_l)^\beta+h.c.\label{eq:effl}
\end{equation}
where $z_{\alpha\beta}$ are couplings, and $i,j,k,l$ are summed over
1,2 (component in the $SU(2)_1$ representation).
Note that (\ref{eq:effl}) violates lepton flavor number by two
units, and preserves the $SU(3)_2$ global
symmetry in the lower $3\times 3$ block of $\Sigma$, 
while breaking the $SU(3)_1$ global
symmetry in the upper $3\times 3$ block of $\Sigma$.
We expand eq.~(\ref{eq:effl}) to leading order in powers of $1/f$:
\begin{eqnarray} 
{\cal L}_{LFV}=&  & -z_{\alpha\beta}\bigg[\bar{\nu^c}_\alpha \nu_\beta
\bigg(2i\phi^0+\frac{1}{f}h^0h^0\bigg) \,-\,
(\bar{\nu^c}_\alpha e_\beta +\bar {e^c}_\alpha \nu_\beta)
\bigg(\sqrt{2}i\phi^+ +\frac{1}{f}h^+h^0\bigg)\nonumber \\
& &\qquad+\bar {e^c}_\alpha e_\beta 
\bigg(2i\phi^{++}+\frac{1}{f}h^+h^+\bigg) \bigg]
+ h.c.
\end{eqnarray}  
Note that the operators are associated with the triplet appear at dimension 4 
while operators associated with the doublet appear at dimension 5.

After $h$ and $\phi$ develop {\it vev}'s, mixing occurs between
them (See the details in Appendix.). The low energy effective operator
is given in terms of the mass eigenstates of the higgs and the
longitudinal components of the gauge fields
\begin{eqnarray} 
&  & -z_{\alpha\beta}\bigg\{2\bar{\nu^c}_\alpha \nu_\beta 
\bigg(v'+\frac{v^2}{4f}\bigg)
-\frac{\bar{\nu^c}_\alpha e_\beta +\bar {e^c}_\alpha \nu_\beta}{\sqrt{2}}
\bigg[ 2\bigg(1-\frac{v'}{f}
-\frac{2v'^2}{v^2}\bigg)\Phi^+
+\frac{4}{f}\bigg(v'+\frac{v^2}{4f}\bigg) G^+\bigg] \nonumber\\ 
& & \,+\,\bar {e^c}_\alpha e_\beta \bigg[
2\Phi^{++}+\frac{1}{f}\bigg[\bigg(1-4\frac{v'^2}{v^2}\bigg) G^+G^+
-4\frac{v'}{v}G^+\Phi^+ + 4\frac{v'^2}{v^2}\Phi^+\Phi^+
\bigg]\bigg]\biggr\} + h.c.\label{eq:long}
\end{eqnarray} 
where $\Phi^+$ and $\Phi^{++}$ are the singly charged and doubly charged
scalars, and $G^+$ is the Goldstone boson that is eaten by the $W^+$
boson, giving it a mass.
Note that the $\bar {\nu^c} \nu$ associated term gives Majorana masses to neutrinos,
so the neutrino mass matrix is then given by
\begin{equation}
[m_\nu]_{\alpha\beta} =2z_{\alpha\beta} 
\bigg(v'+\frac{v^2}{4f}\bigg)\label{eq:nmass}.
\end{equation}

Lightness of neutrinos arises from $z_{\alpha\beta}$.
As shown in Eq.~(\ref{eq:nmass}), the $h$'s {\it vev}
acts like a $\phi$'s {\it vev} at order $v^2/f$.
There is a relation between the two {\it vev}'s by demanding 
the Higgs triplet mass squared to be positive, $v'\lesssim v^2/4f$. 
Furthermore, the current experimental limits on the $\rho$ parameter
lead to more stringent constraint on the Higgs triplet {\it vev},
$v'\lesssim \frac{1}{10} \frac{v^2}{f}$ \cite{Chen:2003fm}.
As a consequence, the Higgs doublet contribution to neutrino mass
is larger than the HIggs triplet contribution: 
$[m_\nu]_{\alpha\beta}\lesssim z_{\alpha\beta} v^2/f$.

The absolute scale of neutrino masses is not determined by the
neutrino oscillations, but can be determined by the observation of the
end-point part of the electron spectrum of Tritium $\beta$-decay,
the observation of large-scale structures in the early universe,
and the dectection of the neutrinoless double beta decay.
From these observations, one can set the upper limit on the heaviest
neutrino mass at  the 0.1$\,\sim\,$1 eV scale.
Then the upper bound of the coupling constants is estimated by
\begin{equation}
|z_{\alpha\beta}|\lesssim 10^{-12}\bigg[\frac{f}{1\,\mbox{TeV}}\bigg].
\label{eq:zmax}
\end{equation}
It is such a tiny number that one may raise a question on its origin. 
In the next section, we will study the composite little Higgs model to 
find a successful mechanism for the tiny couplings.

\section{Composite Little Higgs Model}
\label{sec:origin}

\begin{figure}[htb]
\centerline{\includegraphics[width=9cm]{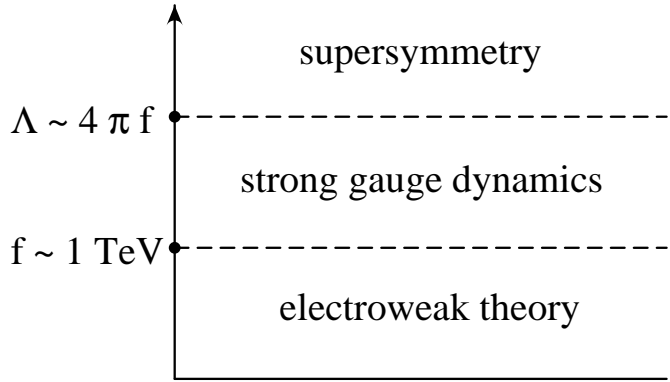}}
{\caption[1]{\label{fig:escale} Energy scale in the composite little
Higgs model}}
\end{figure}
%\FIGURE[l]{{escale.ps}\caption{???}}
The idea of composite Higgs was first introduced by Dugan, Georgi,
and Kaplan \cite{Georgi:1984af}. The composite little Higgs model was
first introduced as a UV completion of the $SU(5)/SO(5)$ little Higgs
model with the $SU(2)_1\times SU(2)_2\times U(1)_Y$ gauge group.
The NGBs arising from the $SU(5)/SO(5)$ global symmetry breaking
are fermion condensation through a strong $SO(7)$ gauge group,
called Ultra-color.  The strong gauge dynamics are merged into
supersymmetry so that the number of supermultiplets in the CLHM
is larger than that in the MSSM. The relevant energy scales in the
composite little Higgs model are shown in Fig. \ref{fig:escale}. Note 
that the UV cutoff scale of the $SU(5)/SO(5)$ little Higgs model is
the same as the supersymmetry breaking scale. The quantum
numbers of matter superfields in the theory are listed in Table 1. 

\TABULAR[l]{|c|ccccc|} 
{\hline
  & $SO(7)$ & $SU(3)_c$ & $SU(2)_2$ & $SU(2)_1$ & $U(1)_Y$  \\ \hline
$L$ & 1 & 1 & 1 & 2 & -1/2 \\
$E^c$ & 1 & 1 & 1 & 1 & 1\\
$Q$ & 1 & 3 & 1 & 2 & 1/6 \\
$U^c$ & 1 & $\bar 3$ & 1 & 1 & -2/3 \\
$D^c$ & 1 & $\bar 3$ & 1 & 1 & 1/3 \\
$\Phi_{\bar 3}$ &  7 & $\bar 3$ & 1 & 1 & -2/3 \\
$\Phi_3$ &  7 & 3 &1 & 1 & 2/3 \\
$\Phi_{2'}$ &  7 & 1 & 2 & 1 & -1/2 \\
$\Phi_2$ &  7 & 1 &1 & 2 & 1/2 \\
$\Phi_0$ & 7 & 1  & 1 & 1 & 0\\
$Y$ & 1 & $\bar 3$ &1 & 2 & -7/6 \\
$Y^\prime$ & 1 & 3 & 2 & 1 & 7/6\\
$H_u$ & 1 & 1 &1 & 2 & 1/2 \\
$H_d$ & 1 & 1 &1 & 2 & -1/2 \\  \hline}
{quantum numbers of chiral superfields}

All the SM particles are neutral under Ultra-color. The chiral superfields
$\Phi_{0,2,2',3,\bar 3}$ are charged under Ultra-color, as
technifermions in the technicolor model. 
$H_u$ and $H_d$ are up-type and down-type Higgs
superfields, respectively, as in the MSSM. The scalar components of
the Higgs fields are irrelevant to the SM Higgs. $Y$ and $Y^\prime$
are introduced to evade the gauge anomalies in the theory.
Their interactions with other supermultiplets are either suppressed 
by their large masses or forbidden by some unknown
global symmetry. For example, a renormalizable superpotential of the
form $QYE^c$ is highly supressed by proton decay.

Let us review the $SU(5)/SO(5)$ symmetry breaking
in detail. The (approximate) $SU(5)$ global symmetry of the
little Higgs model acts on the fields $\Phi_{2,2',0}$.
When the Ultra-color forces become strong at the scale
$\Lambda\sim4\pi f$,
bilinears of Ultra fermions $\tilde\phi_{0,2,2^\prime}$
condense~\cite{Arkani-Hamed:2002qy} as
\begin{equation}
\langle \tilde\phi_{\rho} \tilde \phi_{\sigma} \rangle \approx
4\pi f^3 
S_{\rho\sigma}.
\end{equation}
By comparing hypercharge of bilinears of the condensates
with that of the $\Sigma$ field in the littlest Higgs model, one can see the
correspondence of the Ultrafermion condensates in the composite little
Higgs model to the Higgs fields in the littlest Higgs model. See the
Table. 2.

\TABULAR[l] {|c|c|}{\hline
compostie little Higgs &  littlest Higgs \\ \hline
$S_{22}$ & $\phi$ \\  $S_{2'2'}$ & $\phi^\dagger$ \\
$S_{02}$ & $h$  \\ $S_{02'}$ & $h^\dagger$ \\ \hline}
{Correspondence of Ultrafermion condensates in the composite
little Higgs model to the Higgs fields in the littlest Higgs model} 

We now describe the origin of the Yukawa interactions for 
the {\it light} quarks\footnote{A quark that is lighter than the top
quark.} and leptons. For leptons,
the Yukawa couplings emerge from (\ref{eq:lepmass}), and 
the charged leptons acquire masses from the {\it vev} of the composite
Higgs doublets $S_{02},S_{02'}$. For down type quarks, the Yukawa 
couplings are given in the same fashion as in the leptons. 
In contrast, the Yukawa interactions for up-type quarks are obtained
in slightly different ways: (i)  $\Sigma$ is replaced
by its complex conjugate, $\Sigma^\ast$, and  (ii) the third
generation mixes with vector-like fermions with a TeV mass scale,
and the top quark becomes much heavier than other SM quarks.
\footnote{Heavy composite spin 1/2 fermions are fermion
condensate of the form $\tilde \phi_\rho\tilde \phi_\sigma \lambda$,
where either $\rho$ or $\sigma$ is 3 or $\bar 3$, and $\lambda$
is a gaugino of Ultra-color. Among them charge
2/3 vector-like fermion mixes with the fundamental quark in the third generation.}
In order to introduce light quark and lepton masses in a renormalizable theory, 
we couple the quarks and leptons to the Higgs superfields $H_u$ and $H_d$
in the same fashion as in the MSSM. 

We now describe Yukawa couplings for the light quarks and leptons.
The Yukawa couplings are governed by the the superpotential
\begin{equation}
W=\mu H_uH_d+ \lambda^u_{\alpha\beta}H_uQ_\alpha U^c_\beta
+ \lambda^d_{\alpha\beta}H_dQ_\alpha D^c_\beta
+ \lambda^e_{\alpha\beta}H_dL_\alpha E^c_\beta
+\omega H_d \Phi_2\Phi_0,\label{eq:sumass}
\end{equation}
and soft supersymmetry breaking scalar terms
\begin{equation}
V_{soft}=-(\mu B H_uH_d+h.c.)+M_u^2|H_u|^2+M_d^2|H_d|^2+
M^2_0 |\Phi_0|^2+M^2_2|\Phi_2|^{2}+\cdots,
\end{equation}
where  we take $M_{u,d},M_{0,2}$ to be real and positive, and
omit soft-supersymmetry-breaking mass terms of the SM sfermions
and gluino which are nearly irrelevant to flavor problem.
The $B$ parameter is important in that it couples between
$H_{u}$ and $H_{d}$ bosons.
We adjust the over-all phase of $H_u$ and $H_d$
so that $B$ is real and positive. For simplicity, we assume that
all the mass parameters in (\ref{eq:sumass}) are of the scale $\Lambda$: 
\begin{equation}
\{|\mu|, B, M_{u,d},M_{0,2}\}\sim \Lambda\label{eq:subrk}.
\end{equation}
The $\lambda^{u,d,e}$ parameters
are the Yukawa couplings of the (light) up quark,
down quark and lepton superfields at high energy, respectively.
Note that the $\omega$ term is an addition to the general
renormalizable ($R$-parity conserving) superpotential,
and its main role is to link the SM fermions
and the Ultrafermions at low energy via $H_d$.
For simplicity, we take $\omega$ be real and positive by adjusting 
the over-all phase of $\Phi_0$ and $\Phi_2$. 

\begin{figure}[htb]
\centerline{
\hbox{$\begin{array}{ccc}
\includegraphics[width=4.2cm]{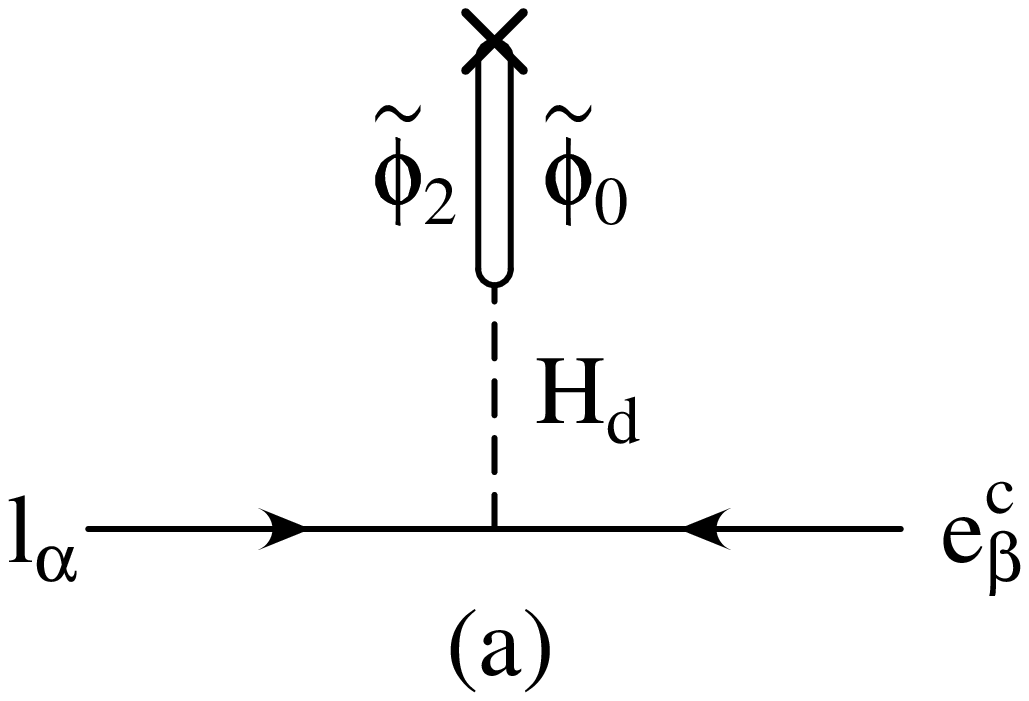} &
\hspace{.1cm}\includegraphics[width=4.2cm]{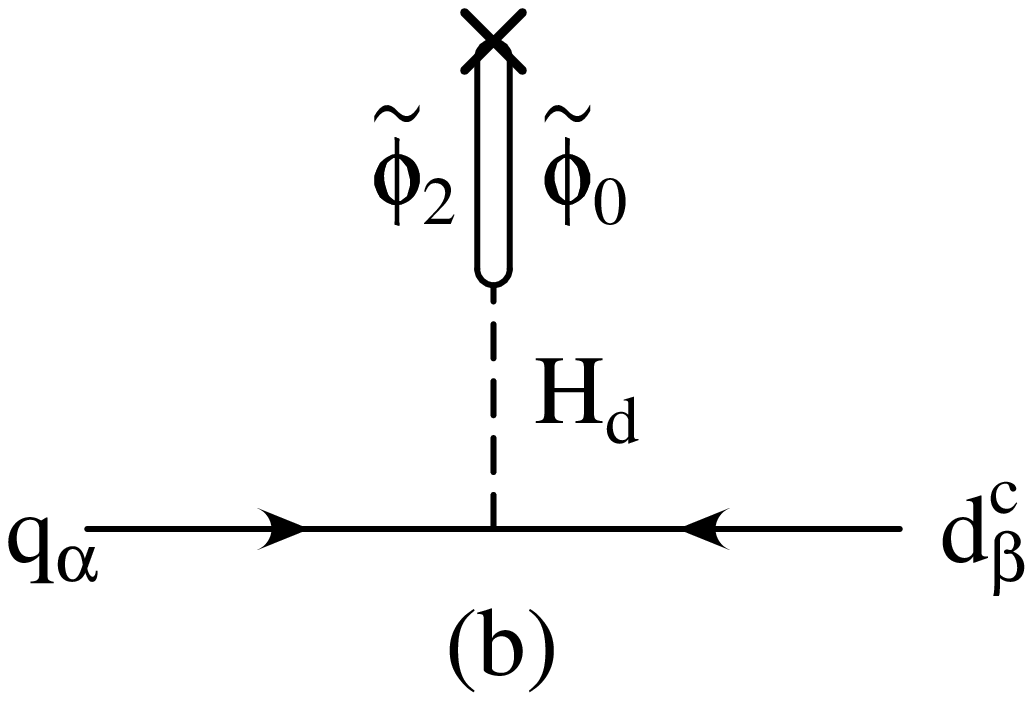} &
\hspace{.1cm}\includegraphics[width=4.2cm]{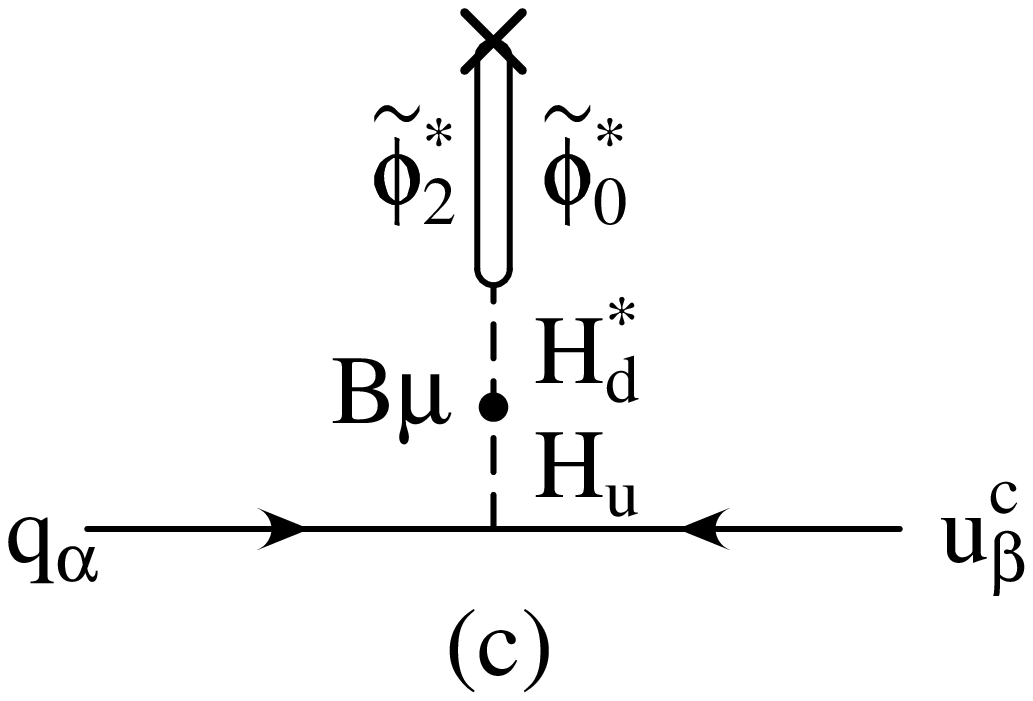} \\ 
\end{array}$}}
{\caption[1]{\label{fig:yukawa}
Diagrams contributing to the Yukawa couplings of (a) charged leptons,
(b) down type quarks, and (c) up type quarks}}
 \end{figure}

Now let us relate the Yukawa couplings above the scale $\Lambda$
to those at the electroweak scale. To do so, we briefly describe
the physics from the top down. At the scale $\Lambda$,
massive superpartners of the SM fields and Ultrascalars decouple.
At the same scale,  the Ultra-color forces become strong so
that the Ultrafermion condensate is triggered, and 
the $SU(5) \to SO(5)$ symmetry breaking occurs. That is,
the composite Higgs triplet and composite Higgs doublet are formed
as a set of NGBs. At the scale $f$,  the composite Higgs triplet gets
a mass of order $gf$. Below the scale $f$,
one gets couplings of the light quarks and leptons to the composite
Higgs doublet to the leading order as follows:
\begin{equation}
y_{\alpha\beta}^uq_\alpha f S_{02} u^c_\beta
+y_{\alpha\beta}^dq_\alpha f S^\ast_{02} d^c_\beta
+y_{\alpha\beta}^e\ell_\alpha f S^\ast_{02} e^c_\beta+h.c.
\end{equation}
where $y^{u,d,e}$ are the Yukawa couplings at low energy.
At the electroweak scale, the composite Higgs doublet acquires a {\it vev}
so that EWSB takes place, and  the light quarks and leptons acquire
masses as proceeding via Ultrascalar exchange, as shown
in Fig.~\ref{fig:yukawa}.\footnote{One may note that
this scenario is very similar to the bosonic technicolor
model (BTM) \cite{Samuel:1990dq}, which merges supersymmetry
and technicolor into a theory to resolve the phenomenological
difficulties that supersymmetry and technicolor separately have.}
Thus one estimate the Yukawa couplings for the light quarks and
leptons at low energy as follows:
\begin{eqnarray}
y_{\alpha\beta}^e &\approx& \frac{\omega\lambda^e_{\alpha\beta}}
{4\pi}\frac{\Lambda^2}{M_d^2+|\mu|^2}\sim
\frac{\omega\lambda^e_{\alpha\beta}}{8\pi},\label{eq:ye}\\
y_{\alpha\beta}^d &\approx& \frac{\omega\lambda^d_{\alpha\beta}}
{4\pi}\frac{\Lambda^2}{M_d^2+|\mu|^2} \sim
\frac{\omega\lambda^d_{\alpha\beta}}{8\pi},\label{eq:byuka}\\
y_{\alpha\beta}^u &\approx& \frac{\omega\lambda^u_{\alpha\beta}}
{4\pi}\frac{\mu B\Lambda^2}{(M_u^2+|\mu|^2)(M_d^2+|\mu|^2)}
\sim\frac{\omega\lambda^u_{\alpha\beta}}{16\pi},
\end{eqnarray}
where we have used (\ref{eq:subrk}).
In order that Yukawa couplings for the up type (light) quark
are comparable in size to those for the down type quark,
one confirm the previous assumption, $B\sim  M_{u,d}$.
Note that the low-energy Yukawa couplings are approximately, by a
factor of $\sim \omega/8\pi$ or $\omega/16\pi$, proportional to the high-energy Yukawa
couplings.

Additional dangerous operators which could violate flavor at low
energy must be proportional to the {\it only} sources of flavor
violation, namely the matrices $\lambda^{u,d,e}$.
For example, the next leading flavor violating diagrams come from
gauge one-loop corrections which are suppressed by a factor of
$\frac{g^2}{16\pi^2}$ compared with the leading diagrams.
One finds that additional dangerous operators which could
violate flavors are  much suppressed, so the low energy Yukawa
matrices are nearly aligned with the matrices $\lambda^{u,d,e}$. 

Given the low-energy Yukawa couplings alone one cannot determine
the value of $\omega$ because the high-energy Yukawa
couplings are still unknown. This provides more uncertainties 
to the parameter spaces in the composite little Higgs model at high energy.
However, $\omega$  is constrained by the largest mass among the light
quarks. The $b$ quark must be taken for the purpose and its Yukawa
coupling is $y^d_{33}\lesssim\sqrt{2}m_b/v$. From (\ref{eq:byuka}),
$\omega$ should be constrained as follows:
\begin{equation}
\omega|\lambda^d_{33}|\approx 0.6\label{eq:olcon}
\end{equation}

As for the SUSY flavor problem, we take the masses of all
the SM superpartners and Ultrascalars  be the UV cutoff scale,
$\Lambda\sim 10$ TeV or above 
so that little or no squark and slepton mass degeneracy is
required at low energy to satisfy the phenomenological 
constraints on CP conserving flavor changing neutral currents (FCNC)
\cite{Nilles:1983ge}. Furthermore, the {\it vev}'s of $H_{u,d}$ in the
composite little Higgs model are much smaller than that in the
MSSM\footnote{We will show it in the next section.}, which gives
an extra suppression in the FCNC processes.
Constraints associated with one-loop contributions to the neutron
electric dipole mement, involving gaugino and squark exchange,
are also satisfied even if CP violating phases are of order one.
Thus, in this scenario, one can naturally resolve the SUSY flavor problem.

However, there are potentially dangerous sources of FCNC due to
a heavy charge 2/3 quark, which is required to give the top quark a large mass.
The CLHM explains the origin of the vector-like quark and its companions
whose masses, if any, lie in the TeV range.
The presence of the heavy vector-like quark induces non-vanishing mixing
angles in the neutral currents sector, and they may give significant
contributions to the FCNC processes at tree level compared to that
from the SM, but all the predictions are beyond the experimental 
sensitivity in near future \cite{Lee:2004me}.

\section{$R$-parity Violations in the Composite Little Miggs Model}

As in the R-parity violating MSSM \cite{Hall:1983id}, the CHLM
does not distinguish between the down-type Higgs superfield $H_d$
and the lepton superfields $L_\alpha$ with respect to
charges under the gauge group. 
As a result, it is convenient to denote these four supermultiplets
by one symbol $L_m\equiv(H_d,L_e,L_\mu,L_\tau)$.
In the followings we use Greek indices for the usual
three dimensional lepton flavor space and Latin indices $m,n$
for the four dimensional extended lepton flavor space.

Now one generalize (\ref{eq:sumass}) to the superpotential
with $R$-parity violations:
\begin{equation}
W=\mu_m L_m H_u + \lambda^u_{\alpha\beta}H_uQ_\alpha U^c_\beta
+ \lambda^d_{m,\alpha\beta}L_mQ_\alpha D^c_\beta
+ \lambda^e_{mn,\alpha}L_mL_n E^c_\alpha
+\omega_m L_m \Phi_0\Phi_2\label{eq:rpvw}
\end{equation}
where $\mu$ and $\omega$ in  (\ref{eq:sumass}) are now extended
to four-component vectors $\mu_m=(\mu,\mu_\alpha)$ and
$\omega_m\equiv(\omega,\omega_\alpha)$ in the extended lepton
flavor space, respectively.
$\lambda^d_{m,\alpha\beta}$ is a vector in a similar fashion
while $\lambda^e_{mn,\alpha}$ becomes a antisymmetric tensor
under the intercahnge of the indices $m,n$.
By adjusting the phases of the lepton superfields $L_\alpha$
one can make either $\mu_\alpha$ or $\omega_\alpha$ positive and real. 
In what follows we take $\omega_\alpha$ be real and positive.
Assuming that the effects of $R$-parity violation is quite small,
one expect that $|\mu_\alpha|\ll|\mu|$ and $\omega_\alpha\ll\omega$.
Note that the small RPV parameters depend on the basis choice
for these superfields although physical observables
are independent of the choice of basis.
In addition, we include soft supersymmetry breaking potential
\begin{equation}
V_{soft}=-(\mu B_m L_mH_u+h.c.)+ M^2_u |H_u|^2
+(M^2_{\tilde L})_{mn} \tilde L_m^{i\ast} \tilde L_n^i 
+(M^2_{\tilde E})_{\alpha\beta} \tilde E_\alpha^\ast \tilde E_\beta
+\cdots,
\end{equation}
where the $B$ parameter in the CLHM is now extended to a four component,
$B_m\equiv(B,B_\alpha)$, with $|B_\alpha| \ll B$ due to small
$R$-parity violation. $B,M_u,M_{\tilde L}$,
$[(M_{\tilde L})_{00}\equiv M_d], M_{\tilde E}$ are all of order $\Lambda$.
Further, we omit other RPV soft-supersymmetry-breaking potentials
which appear in the MSSM.

As mentioned before, at the scale $f$ the composite Higgs doublet
drives small {\it vev}'s of $H_{u,d}$ so that the $W$ and $Z$
acquire small mass contributions from the {\it vev}'s of $H_{u,d}$.
In contrast, above the scale $\Lambda$ the {\it vev}  of composite Higgss
doublet has not turned on yet, so the {\it vev}'s of $H_{u,d}$ is expected
to be zero. We drive the argument by considering the neutral scalar potential
for sneutrinos, $H_u$ and $H_d$.
The contribution of the neutral scalar fields to the scalar potential is given by
\begin{eqnarray}
V_{neutral}&=&(M_u^2+|\hat\mu|^2)|H_u|^2
+ [(M^2_{\tilde L})_{mn}
+\mu_m \mu_n^\ast] \tilde \nu_m \tilde \nu_n^\ast \nonumber \\
&& -(\mu B_m \tilde \nu_m H_u+\mu B_m^\ast \tilde \nu^\ast_m H_u^\ast)
+\frac{1}{8}(g_1^2+g_Y^2)[|H_u|^2-|\tilde \nu_m|^2]^2\label{eq:vne}
\end{eqnarray}
where $\tilde \nu_m\equiv (H_d, \tilde \nu_\alpha)$, 
$|\hat \mu|^2=\sum_m |\mu_m|^2$, and $g_1$ and $g_Y$ are the gauge
couplings in $SU(2)_1$ and $U(1)_Y$ respectively.
The {\it vev}'s for the neutral scalars denoted by
$\langle H_u\rangle\equiv\frac{v_u}{\sqrt{2}}$
and $\langle \tilde \nu_m \rangle \equiv \frac{v_m}{\sqrt{2}}$,
are determined by the following minimization conditions
\begin{eqnarray}
(M_u^2+|\hat \mu|^2) v^\ast_u &=& \mu B_m v_m
-\frac{1}{8}(g_1^2+g_Y^2)(|v_u|^2-|v_d|^2)v^\ast_u \label{eq:cmin1}\\ 
((M^2_{\tilde L})_{mn}+\mu_m \mu_n^\ast ) v_n^\ast&=& \mu B_m v_u
+\frac{1}{8}(g_1^2+g_Y^2)(|v_u|^2-|v_d|^2)v^\ast_m \label{eq:cmin2}
\end{eqnarray}
with $|v_d|^2\equiv \sum_m |v_m|^2$.
Up to this point, there is no preferred direction in the extended lepton
flavor space so that we can choose, for convenience, a basis where
all the sneutrino {\it vev}'s vanish such that $v_m=(v_d,0,0,0)$.
\footnote{We use the basis-independent parameters constructed in
and write the neutrino mass matrix in terms of various PRV parameters
$\delta^\alpha_\mu,\delta^\alpha_B,\delta_{\lambda^d}^{m\alpha\beta}$
and $\delta_{\lambda^e}^{mn\alpha}$~\cite{Davidson:2000uc}.
In the basis where the sneutrino {\it vev}'s are zero,
the parameters correspond to $\mu_\alpha/|\mu|,B_\alpha/|B|,
\lambda_{m\alpha\beta}^d$ and $\lambda_{mn\alpha}^e$,
respectively.}
In this basis, without changing the over-all phase of $H_{u,d}$
we adjust their relative phases so that $v_u$ is real. Then
eqs. (\ref{eq:cmin1}) and (\ref{eq:cmin2}) show that $v_d$ is also real
so that these equations become
\begin{eqnarray}
(M_u^2+|\hat \mu|^2) v_u &=& \mu B v_d
-\frac{1}{8}(g_1^2+g_Y^2)(v_u^2-v_d^2)v_u \label{eq:mmx1}, \\ 
(M^2_d+|\mu|^2) v_d &=& \mu B v_u
+\frac{1}{8}(g_1^2+g_Y^2)(v_u^2-v_d^2)v_d \label{eq:mmx2}, \\
((M^2_{\tilde L})_{\alpha 0}+\mu_\alpha \mu^\ast ) v_d &=&
\mu B_\alpha v_u \label{eq:mmx3}.
\end{eqnarray}
To satisfiy the assumption (\ref{eq:subrk}) with the {\it vev}'s of
$H_{u,d}$, $v_{u,d}\ll f$,
there is {\it only} one solution for the Higgs {\it vev}'s, $v_u=v_d=0$.
That is,  the gauge group is unbroken above the scale $\Lambda$.
However, the $SU(2)_1\times SU(2)_2$ gauge groups are broken 
into its diagonal subgroup $SU(2)_L$ at the scale $f$ and
the broken gauge bosons acquire mass of order $gf$. 
Furthermore, at the electroweak scale the SM gauge groups are
broken and the $W$ and $Z$ bosons acquire masses.

The contributions to the $W$ and $Z$ masses arise after the {\it vev}'s
of the Ultrafermion condensate are turned on.
The $W$ and $Z$ bosons acquire masses from different two sources; 
the leading contribution comes directly from the {\it vev}  of the
composite Higgs doublet, $v$ and the other indirectly from the {\it vev}
of $H_{u,d}$ which are driven by the composite Higgs. But the latter is
much smaller than the former. The ratio of the {\it vev}  of $H_d$ 
to the {\it vev}  of the composite Higgs doublet is given by
\begin{equation}
\frac{v_{d}}{v}\approx
\frac{\frac{\omega}{4\pi}\frac{\Lambda^2}{M_d^2+|\mu|^2}
v} {v}
\sim \frac{\omega}{8\pi}\label{eq:twovv}\ll 1,
\end{equation}
where we have assumed $\Lambda\sim M_d$. One can impose
severe contraint on the ratio by precision electroweak data.
The $\omega$ parameter can not be arbitrarily small because it should meet
the condition (\ref{eq:olcon}) as well. By taking $\lambda^d_{33}\sim 1$,
we set $\omega$ in the range of $0.1\lesssim\omega\lesssim1$. For 
$\omega=1$, we estimate the upper limits on the $\lambda^{d,e}$ and 
list them in Table 3.
\TABULAR[l] {|c|c|}{\hline
$\lambda$ & upper limit \\ \hline
$d$ & $1\times 10^{-3}$ \\ 
$s$ & $2\times 10^{-2}$ \\
$b$ & 0.6 \\ \hline
$e$ & $7\times 10^{-5}$ \\
$\mu$ & $1\times 10^{-2}$ \\
$\tau$ & $1\times 10^{-2}$ \\ \hline}
{Upper limits on the elements on the Yukawa matrices at high energy
for $\omega=1$}

\section{Contributions to the Neutrino Masses}

In the previous section, we have shown that the RPV terms allow
lepton number violation by one unit.
Two of these are taken together to construct the low-energy effective
operators, which violate lepton number by two units.
The neutrino mass matrix arises from both tree- and
loop-level diagrams, as in the RPV MSSM.
Many RPV parameters are involved in these operators.
In the RPV MSSM, there is a region of the RPV parameter spaces
where the bilinear $\mu_\alpha$ term dominantly contributes
to the neutrino mass matrix in order naturally to describe the neutrino
mass hierarchy in neutrino oscillation experiments \cite{Grossman:1997is}:
The bilinear term gives mainly  the largest neutrino mass and the
one-loop contributions are subordinate, inducing other light
neutrino masses. 

The composite little Higgs model has an additional RPV paramter
compared with the RPV MSSM. This is the $\omega_\alpha$ parameter.
Though the $\omega_\alpha$-term is trilinear it behaves
as a bilinear term below the scale $f$. Among all the contributions
to the neutrino mass matrix,
only the effective opeartor consisting of the $\omega_\alpha$ term
appears in the $SU(5)/SO(5)$ little Higgs model. In the absence
of the $\mu_\alpha$ contribution to the neutrino mass, the
$\omega_\alpha$ term can, in turn, provide another
economical framework for the solution of neutrino
mass hierarchy. In the following we quantitatively  analyze the
tree- level contributions while we qualitatively comment on one loop-level
contributions to the neutrino mass.

\subsection{Tree level $(\mu\mu)$ contribution}

\begin{figure}[htb]
\centerline{\hbox{\includegraphics[width=6cm]{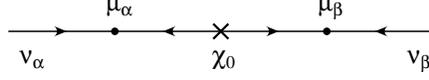}}}
{\caption[1]{\label{fig:uvtree} Tree level neutrino mass in the mass
insertion approximation. A blob represents mixing between the
neutrino and the up-type Higgsinos/gauginos.
The cross on the neutralino propagator signifies a Majorana
mass term for the neutrino.}}
\end{figure}

In the previous section, we have shown that $\mu$ extended to
a four-vector, $\mu_m$, and this admits bilinear RPV interactions
between leptons and Higgsinos,
\begin{equation}
\mu_\alpha L_\alpha H_u + h.c.
\end{equation}
Two of these interactions are joined together and then
$\Delta L=2$ LFV interactions are induced through mixing with the
neutralinos as shown in Fig. \ref{fig:uvtree}. 

As in a generic RPV MSSM,
the $7\times7$ gaugino-Higgsino-neutrino mass matrix
in a basis spanned by the two neutral gauginos, the Higgisinos,
and three generations of neutrinos
is non-diagonal:
\begin{equation}
M^{(n)}=\left(\begin{array}{ccccccc}
M_1 & 0 & \frac{1}{2}g_Y v_u & -\frac{1}{2}g_Y v_d &0&0&0\\
0& M_2 & -\frac{1}{2}g_1 v_u & \frac{1}{2}g_1 v_d &0&0&0\\
\frac{1}{2}g_Y v_u & -\frac{1}{2}g_1 v_u &0&
-\mu&-\mu_1&-\mu_2&-\mu_3\\
-\frac{1}{2}g_Y v_d & \frac{1}{2}g_1 v_d &-\mu& 0&0&0&0\\
0&0&-\mu_1&0&0&0&0\\
0&0&-\mu_2&0&0&0&0\\
0&0&-\mu_3&0&0&0&0 
\end{array}\right)
\end{equation}
where $M_{1,2}$ are the gaugino mass parameters, and
mass hierarchy among the parameters is 
$M_{1,2}\sim \mu\gg v\gg v_d$.

Assuming that all the masses of gauginos and Higgisinos are of order
10 TeV or higer,  one can integrate out all of them at low energy.
Thus the mass matrix for remaining neutrinos is estimated by
\begin{equation}
[m_\nu]^{(\mu\mu)}_{\alpha\beta} \approx
\bigg[\frac{(g_1v_d)^2}{4M_2}+\frac{(g_Y v_d)^2}{4M_1}\bigg]
\bigg(\frac{\mu_\alpha \mu_\beta}{\mu^2}\bigg).
\end{equation}
This constitutes a rank 1 mass matrix, leading to only
one nonzero mass eigenvalue which is presumably $m_3$:
\begin{equation}
m_3\approx
\bigg[\frac{(g_1v_d)^2}{4M_2}+\frac{(g_Y v_d)^2}{4M_1}\bigg]
\bigg(\frac{\sum_\alpha |\mu_\alpha|^2}{\mu^2}\bigg).
\end{equation}
Note that it is suppressed by a huge scale difference between 
Majorana gaugino mass and the down type Higgs {\it vev}. In contrast,
they are same order in the RPV MSSM. For $m_3\lesssim 0.1$ eV,
$|\mu|\sim M_{1,2}\sim 10$ TeV, and $g_1 v_d\sim g_Y
v_d\lesssim10$ GeV,\footnote{Here we set $\omega\approx 1$ 
and $g_1\lesssim 1$.} the upper bound of the RPV parameter
$\delta_\mu^\alpha(\equiv\mu_\alpha/|\mu|)$ is estimated by
\begin{equation}
\delta_\mu^\alpha\lesssim 10^{-4}
\bigg(\frac{10\mbox{ GeV}}{g_1v_d}\bigg)
\bigg(\frac{M_{1,2}}{10\mbox{ TeV}}\bigg)^{1/2}.
\end{equation}
Note that upper bound of $\delta^\mu_\alpha$ is sensitive to
the down-type Higgs {\it vev} : The smaller the down type Higgs
{\it vev} $v_d$ is, the larger the RPV parameter ratio
$\delta^\mu_\alpha$ is. However, $v_d$ is, as described in
(\ref{eq:twovv}), not arbitrarily small. On the contrary,
large value of $\delta^\mu_\alpha$ is not preferred due
to the assumption of small RPV parameter in the beginning.

\subsection{One-loop neutrino masses}

Combination of two among $\Delta L=1$ LFV interactions in the
superpotential and in the soft supersymmetry breaking scalar mass
terms gives rise to a one-loop induced neutrino mass with the down
squark (slepton) and antisquark(antislepton) pairs being exchanged
in the loops along with their ordinary partners, just as in the RPV MSSM
\cite{Hall:1983id,Grossman:1997is,Rakshit:2004rj}.
Any one-loop contribution can not induce the LFV operator in the
$SU(5)/SO(5)$ little Higgs model due to no couplings to the Ultrafermions.
The squarks (sleptons) in the composite little Higgs model have larger masses
of order 10 TeV, and the upper bounds on the RPV parameters, in
general, are less than those in the RPV MSSM.

In the following, we summarize the approximate expressions of various
one-loop contributions to neutrino mass matrix
\cite{Grossman:2003gq,Rakshit:2004rj}:
\begin{equation}
[m_\nu]^{(\mu \lambda^d)}_{\alpha\beta} \sim 
\sum_{\gamma}\frac{3g}{16\pi^2}m_{d_\gamma}
\frac{\mu_\alpha \lambda^d_{\beta,\gamma\gamma}
+\mu_\beta \lambda^d_{\alpha,\gamma\gamma}}{\Lambda},
\end{equation}
\begin{equation}
[m_\nu]^{(\lambda^d\lambda^d)}_{\alpha\beta} \sim 
\sum_{\gamma,\delta}
\frac{3}{8\pi^2}\lambda^d_{\alpha,\gamma\delta}
\lambda^d_{\beta,\delta\gamma}
\frac{m_{d_\gamma} m_{d_\delta}}{\Lambda},
\end{equation}
\begin{equation}
[m_\nu]^{(\lambda^e\lambda^e)}_{\alpha\beta} \sim 
\frac{1}{8\pi^2}\lambda^e_{\alpha,\gamma\delta}
\lambda^e_{\beta,\delta\gamma}
\frac{m_{l_\gamma} m_{l_\delta}}{\Lambda}, 
\end{equation}
\begin{equation}
[m_\nu]^{(BB)}_{\alpha\beta} \sim 
\frac{g^2}{64\pi^2} \frac{B_\alpha B_\beta}{\Lambda}\epsilon,
\end{equation}
where we ignore the $(\mu\lambda^e)$ and $(\lambda^e\lambda^e)$ loop
contributions due to Yukawa suppression of the down type quarks and a
suppression factor, $\epsilon'$  in $(\mu B)$ loops are expected
to be $\lesssim 1$.
The $(\mu B)$ and $(BB)$ loops are least constrained so that
they may give the largest loop contributions. Taking,
$\delta_\mu^\alpha \sim 10^{-4}$, this leads to $\delta^\alpha_B
(\equiv B_\alpha/B)\sim 10^{-7}$.
It is interesting to note that the required sizes of the $\delta$'s
are too small, where as we expect them to be naturally of order one. 
The smallness of $\delta$'s can be understood in the framework
of some horizontal symmetries which are spontaneously broken by {\it
vev}'s of some ``flavoron'' fields, as in the RPV MSSM \cite{Grossman:2003gq}.
In order to maintain neutrino mass hierarchy in a simple framework,
we still take the assumption that one-loop diagrams contribute
insignificantly to the heaviest neutrino mass in the presence of
large tree-level contributions.

\subsection{$(\omega\omega)$ contribution}

Now we focus on the $\omega_{\alpha}$-term in (\ref{eq:rpvw}),
which is a trilinear interaction:
\begin{equation}
\omega_\alpha\epsilon_{ij} L_\alpha^i \Phi_2^j\Phi_0+ h.c.
\end{equation}
where we write down explicitly dependence of the gauge indices $i,j=1,2$. 
\begin{figure}[htb]
\centerline{
\hbox{$\begin{array}{cc}
\includegraphics[width=4.5cm]{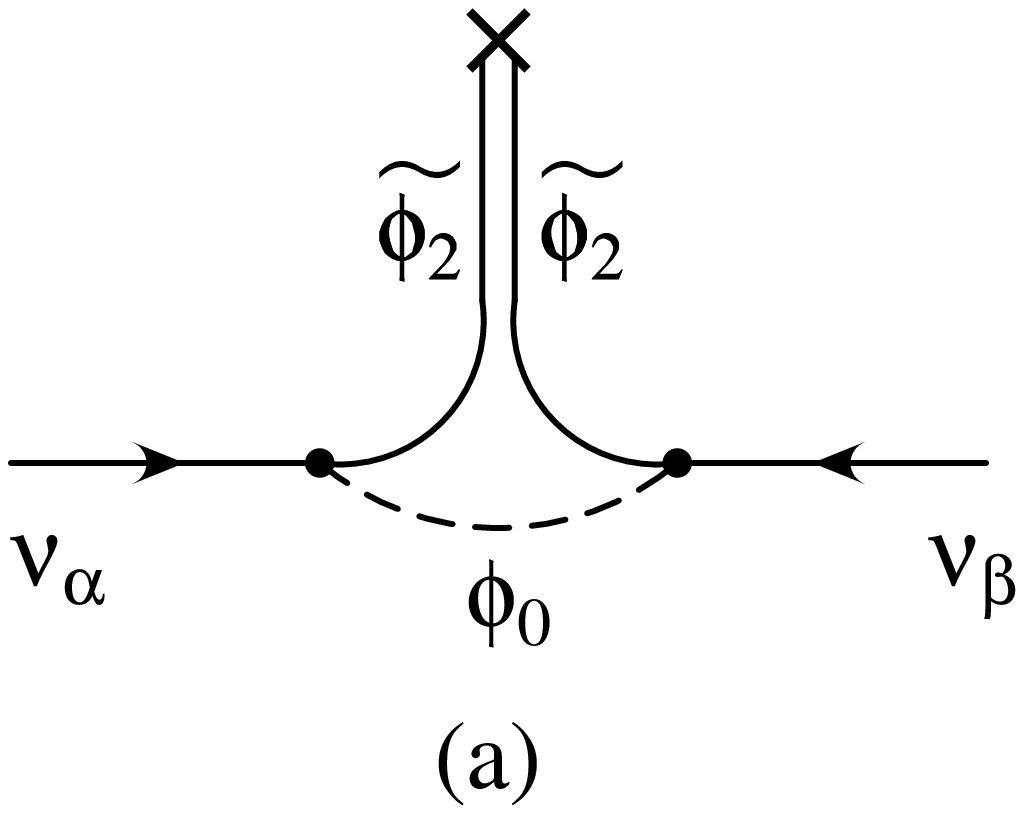} &
\hspace{1cm}\includegraphics[width=4.5cm]{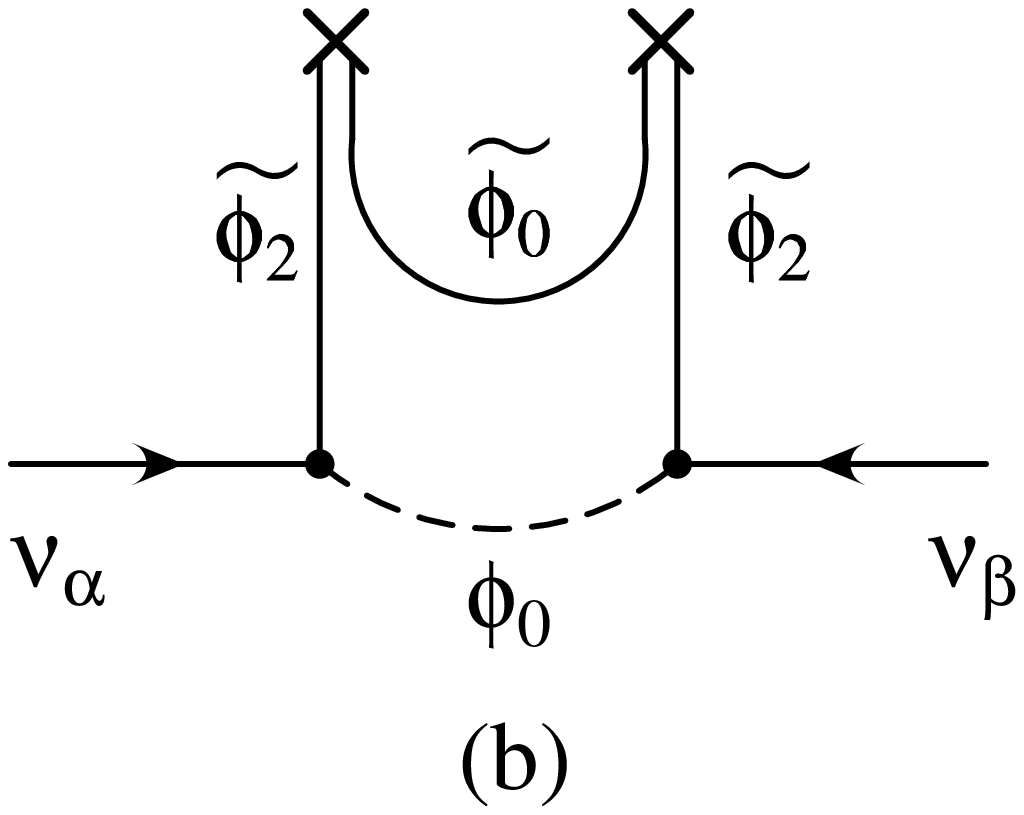} \\
\end{array}$}}
{\caption[1]{\label{fig:uvsusy} Diagrams generating the LFV operator,
(\ref{eq:effl}). (a) represents interaction with the composite Higgs triplet.
(b) represents interacton with the composite Higgs doublet.}}
\end{figure}
Two of these interactions are combined together,
and $\Delta L=2$ LFV interactions are induced below
the scale $\Lambda$ through the Ultrafermions condensation.
The two diagrams shown in Fig.~\ref{fig:uvsusy} 
give rise to neutrino mass matrix below the scale $f$.
Taking the mass of the Ultrascalar $\phi_0$ be of the scale  
$\Lambda\gtrsim 10$ TeV as in Ref.~\cite{Katz:2003sn},
one has the neutrino mass matrix as follows,
\begin{equation}
[m_\nu]^{(\omega\omega)}_{\alpha\beta} \approx
\frac{\omega_\alpha \omega_\beta}{4\pi}
\frac{\Lambda^2}{M_0^2}\bigg(v'+\frac{v^2}{4f}\bigg)
\approx \frac{\omega_\alpha \omega_\beta}{4\pi}
\bigg(v'+\frac{v^2}{4f}\bigg)\label{eq:nuom}.
\end{equation}
This also constitutes a rank 1 mass matrix, leading to {\it only}
one nonzero eigenvalue,
\begin{equation}
m_3\approx \frac{1}{4\pi}\bigg(v^\prime+\frac{v^2}{4f}\bigg)
(\omega_1^2+\omega_2^2+\omega_3^2)\label{eq:m3m}.
\end{equation}
Comparing it with the definition of $z_{\alpha\beta}$
in (\ref{eq:long}) one write down the coupling constant
in terms of $\omega_\alpha$
\begin{equation}
z_{\alpha\beta}=\frac{\omega_\alpha \omega_\beta}{8\pi}.
\label{eq:bypro}
\end{equation}

\section{Leading $(\omega\omega)$ Contribution in Neutrino Masses}

In the present section, we consider the region in the RPV parameter
spaces where the $(\omega\omega)$ contribution to the neutrino
mass dominates over all the other contributions.
This is an interesting region because the neutrino mass is
almost irrelevant to the UV completions of the $SU(5)/SO(5)$ little
Higgs model. From  eq. (\ref{eq:nuom}), the neutrino mass matrix
elements depend both on  the ratio of the Ultrascalar mass
to the UV cutoff scale and on the three $\omega_\alpha$'s.
By rescaling the $\omega_{\alpha}$,
$\omega_{\alpha}\rightarrow\frac{M_0}{\Lambda}\omega_{\alpha}$,
one sees the ratio disappear in eq. (\ref{eq:nuom}).
Thus, without knowledge of the Ultrascalar
mass and the scale $\Lambda$ one can analyze the neutrino mixing
angles only with $\omega_\alpha$'s.  

The neutrino $\nu_\alpha$ in the weak eigenstates are mixtures
of the neutrino $\nu_i$ in mass eigenstates with the mass $m_{i} \,(i=1,2,3)$
\begin{equation}
    \nu_\alpha=\sum_{i} U_{\alpha i} \nu_i,
\end{equation}
where $U_{\alpha i}$ is a $3\times 3$ unitary mixing matrix
parameterized in the Kobayashi-Maskawa manner. Then the
diagonalization of the neutrino mass matrix is given by
\begin{equation}
U^\dagger [m_\nu] U= \mbox{diag}[m_1,m_2,m_3],\label{eq:num123}
\end{equation}
where we choose a basis where neutrino mass eigenstates are in
ascending order of mass, $(m_1,m_2,m_3)$.
With that in mind, we analyze the neutrino mixing angles in terms of
$\omega_{\alpha}$'s. One gets a simple relation for $\hat \omega_\alpha
(\equiv\omega_\alpha /\sqrt{\omega_1^2 +\omega_2^2+\omega_3^2})$ 
from eqs. (\ref{eq:nuom}) and (\ref{eq:num123}): 

\begin{equation}
\hat \omega_\alpha\hat \omega_\beta
=U\mbox{diag}\,(0,0,1)U^T
\end{equation}
where the mixing matrix $U$ is parameterized as
\begin{equation}
U = \left( \begin{array}{ccc}
c_{12} c_{13} &
s_{12} c_{13} &
s_{13} \\
-s_{12} c_{23} - c_{12} s_{23} s_{13}&
c_{12} c_{23} - s_{12} s_{23} s_{13}&
s_{23} c_{13} \\
s_{12} s_{23} - c_{12} c_{23} s_{13} &
-c_{12} s_{23} + s_{12} c_{23} s_{13} &
 c_{23} c_{13}
\end{array} \right).
\label{Ubar}
\end{equation}
Here we ignore CP phase because the three $\omega_\alpha$ are, by
definition, all real parameters.
Then we identify $\hat\omega_\alpha$ with the mixing angles:
\begin{equation}
\hat\omega_1=s_{13},\quad
\hat\omega_2=s_{23}c_{13},\quad
\hat \omega_3=c_{23}c_{13}.
\end{equation}
With the experimental data from the recent neutrino oscillation
experiments \cite{Giacomelli:2004va},
\begin{eqnarray}
  &&\Delta m^2_{32} =2.2\times 10^{-3}\mbox{ eV}^2, \qquad
  \Delta m^2_{21} =8.1\times 10^{-5}\mbox{ eV}^2, \\ \nonumber
  &&\sin^2\theta_{23}=0.50,\qquad
  \sin^2\theta_{12}=0.30,\qquad
  \sin^2\theta_{13}<0.068,
\end{eqnarray}
one obtains the approximate values of
$\hat\omega_{\alpha}$'s: 
\begin{equation}
\hat\omega_1\approx 0,\qquad 
\hat\omega_2=\hat\omega_3\approx \frac{1}{\sqrt{2}},
\end{equation}
which implies that\footnote{we set $f=1$ TeV, $v'\ll\frac{v^2}{4f}$.}
\begin{equation}
    \omega_2\approx \omega_3\approx 4\times 10^{-6},\qquad
    \omega_1\ll\omega_2.
\end{equation}
This leads to $\delta^\alpha_\omega(\equiv\omega_\alpha/\omega)
\sim 10^{-6}$.

There are special regions of the parameter spaces where $\mu_{\alpha}$
or $B_\alpha$ are parallel to $\omega_{\alpha}$. In these regions, 
neutrino mass hierarchy is still maintained.
For more complicated situations, {\it i.e.}  the $\mu_\alpha$
tree-level diagram or one loop-level diagrams are competitive in
amplitude with the $\omega_\alpha$ diagram, one can consider
degenerate neutrino mass patterns.

\section{Conclusion}
 
The $SU(5)/SO(5)$ little Higgs model describes the SM Higgs
as a subset of pNGBs, which implies the UV cutoff is typically
in the range of $10\lesssim\Lambda\lesssim100$ TeV. We have
constructed a low-energy effective operator which describes lepton
flavor violation in the $SU(5)/SO(5)$ little Higgs model. After
acquiring the {\it vev}'s, the Higgs doublet and the Higgs triplet
give rise to a neutrino mass matrix. The origin of the small
neutrino mass is not understood in the model itself.

The composite little Higgs model is a UV completions of the
$SU(5)/SO(5)$ little higgs model, which combines strong gauge
dynamics with supersymmetry. We have further worked out flavor
physics of the composite little Higgs model, in particular, the fermion mass
generation via four-fermi interaction between ordinary fermions
and Ultra fermions. The mediators of the four-fermi interaction are
the Ultrascalars with mass of order 10 TeV or higer, which naturally 
explains the suppression of FCNC at low energy. 

$R$-parity violation is incorporated into the composite little Higgs model
such that lepton flavor violation and neutrino mass are naturally
explained. We have identified various lepton flavor violating operators
with $R$-parity violating interactions, and have studied the region of
the RPV parameter spaces by contraints on the upper limits on the
heaviest neutrino mass. In particular, we have identified the presence
of the low-energy LFV effective operator in the $SU(5)/SO(5)$ 
little Higgs model, and have analyzed a special region of RPV parameter
spaces where $(\omega\omega)$ contribution to neutrino mass matrix
dominantes over other contributions. 

One can further ask questions on lepton violating processes like
$\mu\to e\gamma$ and $\mu\to 3e$ in the composite little Higgs model.
We expect that the prediction of these processes are much lower than
the current experimental bounds because the Ultrascalar mass is
larger than the squark and slepton masses in the MSSM, and further
the relevant couplings $z$'s are very small.

Finally, we would like to comment on the virtue of the composite little
Higgs model. Though the MSSM is the most realistic supersymmetric
model it suffers from a naturalness problem, called the `supersymmetric
little hierarchy problem'.  Little Higgs was introduced as an
alternative to supersymmetry.  Combining little Higgs with
supersymmetry is a possible path to avoid the naturalness  problem.
One can apply both supersymmetry and strong gauge dynamics to find
a UV completions of the other little Higgs models. The first thing to 
do is to find a correct strong guage dynamics which contains the pNGBs 
of the model as fermion condensates.

\section{Acknowledgements}

J.L. would like to thank Ann Nelson for stimulating conversations.
J.L. was partially supported by the DOE under contract
DE-FG03-96-ER40956, and the RRF research Grant from the University of 
Washington.

\section{APPENDIX}

The gauge eigenstates of the Higgs fields $h^+$ and $\phi^+$ can
be written in terms of the mass eigenstates of the Higgs fields
$G^+$ and $\Phi^+$ as follows:

\begin{eqnarray*}
h^0 &=& \big(c_0H-s_0\Phi^0+v\big)\sqrt{2}
+i\big(c_PG^0-s_P\Phi^P)/\sqrt{2}, \\
\phi^0 &=&\big(s_PH+c_P\Phi^0)\sqrt{2}
-i\big(s_0H+c_P\Phi^0+\sqrt{2}v')/\sqrt{2}, \\
h^+ &=& c_+G^+-s_+\Phi^+, \\
\phi^+&=&\big(s_+G^++c_+\Phi^+\big)/i, \\
\phi^{++} &=& \Phi^{++}/i
\end{eqnarray*}

We use the following notation for the physical mass eigenstates:
$H$ and $\Phi^0$ are neutral scalars, $\Phi_P$ is a neutral pseduscalars,
$\Phi^+$ and $\Phi^{++}$ are the charged and doubly charged scalars, and
$G^+$ and $G^0$ are the Goldstone bosons that are eaten by the light 
$W$ and $Z$ bosons, giving them mass. Note that in defing the mass 
eigenstates we have factored out an $i$ from $\phi$.
The mixing angles in the pseudoscalar and singly-charged sectors
are easily extracted in terms of the vacuum expectation values:

\[ s_P=\frac{2\sqrt{2}v'}{\sqrt{v^2+8v'^2}}\simeq \frac{2\sqrt{2}v'}{v},
\quad c_P=\frac{v}{\sqrt{v^2+8v'^2}}\simeq 1-\frac{4v'^2}{v^2}\]                                                                         
\[ s_+=\frac{2v'}{\sqrt{v^2+4v'^2}}\simeq \frac{2v'}{v},
\quad c_+=\frac{v}{\sqrt{v^2+4v'^2}}\simeq 1-\frac{2v'^2}{v^2}\]   

\

Diagonizing the mass terms for the neutral CP-even scalars gives the 
scalars mixing angle $s_0$, $c_0$ to leading order in $v/f$:

\[ s_0 \simeq 2\sqrt{2} \frac{v'}{v}, 
\quad c_0 \simeq 1-4\frac{v'^2}{v^2} \]

\end{document}